\documentclass[12pt]{article}
\usepackage[english]{babel}

\usepackage[letterpaper,top=2cm,bottom=2cm,left=3cm,right=3cm,marginparwidth=1.75cm]{geometry}

\usepackage[colorlinks=true, allcolors=blue]{hyperref}
\usepackage{url}
\usepackage{graphicx}
\usepackage[english]{babel}

\usepackage[style=apa, backend=biber, natbib]{biblatex}
\usepackage{amsmath,amsfonts,amssymb,amsthm,mathtools,mathrsfs}
\usepackage{csquotes}
\usepackage{caption,subcaption}
\usepackage{xcolor}
\usepackage{soul}
\usepackage{float}
\usepackage{multirow}

\begin{document}

\title{Deep Learning for Dynamic NFT Valuation}

\author{Mingxuan He\\
	University of Chicago
    \thanks{mingxuanh@uchicago.edu, Chicago, IL, United States of America}}
\date{\today}

\maketitle

\begin{abstract}
    I study the price dynamics of non-fungible tokens (NFTs) and propose a deep learning framework for dynamic valuation of NFTs. I use data from the Ethereum blockchain and OpenSea to train a deep learning model on historical trades, market trends, and traits/rarity features of Bored Ape Yacht Club NFTs. After hyperparameter tuning, the model is able to predict the price of NFTs with high accuracy. I propose an application framework for this model using zero-knowledge machine learning (zkML) and discuss its potential use cases in the context of decentralized finance (DeFi) applications.\\


\end{abstract} 


\section{Introduction}
\label{sec: introduction}
A non-fungible token (NFT) is a digital asset stored on a blockchain. The uniqueness of each NFT asset is verifiable through the unique identification assigned by the blockchain on which they exist.
NFTs can represent photos, videos, audio, and other types of digital files. NFTs are most commonly bought and sold with cryptocurrencies via digital marketplaces, with the Ethereum blockchain being the most popular market.

Recently, picture NFTs have emerged as a popular investment vehicle. For example, the NFT of a digital artwork by Beeple was sold for \$69 million in March 2021. The NFT market has grown rapidly since 2019, with the total trading volume exceeding \$12 billion in Q1 2022 alone.

Similar to their fungible counterpart (cryptocurrencies), a picture NFT can vary from a few dollars to millions of dollars. The price of a picture NFT is determined by a variety of factors, most notably the possession of ``rare traits". However, similar to the traditional arts market, most NFT prices are also heavily dependent on market taste and sentiment, such as the current popularity of the collection and/or the artist.

In this paper, I apply machine learning methods to value the price of picture NFTs. Specifically, I train models on market prices and rarity data obtained from the Ethereum blockchain. The goal is to estalish a model framework that provides accurate estimates of NFT prices. Such a model has various applications, including but not limited to onchain trading and bidding contracts, risk management for crypto portfolios, and fair valuation of staked/collateralized NFTs for decentralized finance (DeFi) applications. In addition, the model can be used to price large batches of NFTs quickly, which is useful for AI-generated NFTs.

The rest of this paper is structured as follows. Section \ref{sec: lit_review} discusses the existing literature on this topic and the novelty of this paper. Section \ref{sec: data} provides an overview of the data used in this project. Section \ref{sec: method} outlines the general methodology and the machine learning models used in training. Section \ref{sec: results} presents the results. Section \ref{sec: conclusion} draws conclusions.

\section{Literature Review}
\label{sec: lit_review}
The fast-growing market of NFTs has attracted interest from various academic disciplines. In the economics and finance literature, various authors have discussed the price dynamics of the NFT market. A comprehensive work by \citet{nadini2021mapping} mapped out important statistical features of NFT markets, revealing that mean prices, sales per asset, sales per collection all follow power-law distributions. \citet{ante2022non} found that NFT sales are triggered by price shocks in Bitcoin (BTC) and active NFT wallets are reduced by price shocks in  Ether (ETH). Similarly, \citet{dowling2022non} found co-movement between cryptocurrencies and NFTs through wavelet coherence analysis. On the buyer side, \citet{kong2021alternative,oh2022investor} found that well-connected and experienced investors generally earn higher returns on NFT trades. These results support the general belief that the NFT market is not fully efficient, and there is room for improvement in pricing NFTs. 

The literature agrees on that rarity is one of, if not the most important factors in determining the price of NFTs. Using data from the CryptoPunks collection, \citet{kong2021alternative} built a hedonic regression model and highlighted rarity as a key determinant of price premium in the cross-section. \citet{mekacher2022heterogeneous} used a custom-built rarity score and showed that rarer NFTs sell for higher prices, are traded less frequently, guarantee higher returns, and are less risky. Taking a network clustering approach, \citet{nadini2021mapping} extracted vector representations of the visual features of NFTs and analyzed their cosine distance network using the pre-trained convolutional neural network AlexNet. These results inspired the inclusion of trait features and the OpenRarity rarity rank as a key set of features in my model.

In the applied machine learning literature, there have been notable attempts to train predictive models on visual features of NFTs. \citet{nadini2021mapping} built a linear regression model with extracted principal components of the NFT's image, features of the trading network, and market history. \citet{seyhan2023nft} and \citet{costa2023show} used multimodal data including NFT images and text characteristics while excluding financial features in an effort to predict both primary and secondary sale prices. The literature also identified other features such as search trends \citep{jain2022nft,kaneko2021time}, social media influence e.g. Twitter \citep{kapoor2022tweetboost}. My proposed framework contributes to the literature by incorporating a large set of tabular features and build a deep learning model tailored to each collection. This approach is more scalable and can be applied to a wide range of collections for secondary sale price prediction and valuation, especially in fully decentralized applications where image data is costly to store and transfer onchain.

\section{Background \& Data}
\label{sec: data}

\subsection{Institutional background}
The NFT market is an emerging market with unique aspects. In this section, I provide a brief overview of the background of NFT markets, including relevant terminology.

\begin{itemize}

    \item \textbf{NFT Marketplaces:} NFT Marketplaces are platforms built with automated blockchain contracts (known as ``smart contracts'') to facilitate transactions using cryptocurrencies. Most popular marketplaces include OpenSea, Blur, Rarible, SuperRare, etc. 

    \item \textbf{NFT Collections:} Many NFTs are created in collections, in which all NFTs share a common theme. For example, the CryptoPunks collection consists of 10,000 unique pictures of pixelated faces. 

    \item \textbf{Traits:} Each NFT in the collection has a unique combination of traits, including color, background, face/expression, and accessories. The traits typically differ by rarity. For example, there are more than 2,000 CryptoPunks with the trait ``Earring", but only 44 with the trait ``Beanie".
    
    \item \textbf{Rarity Score \& Rarity Rank: } The current industry standard for calculating the rarity of an NFT within a collection is the OpenRarity Standard \footnote{\url{https://www.openrarity.dev/}}, where the rarity of an NFT is evaluated on the rarity of its traits. The calculations for this metric is outlined in Appendix \ref{app: rarity}. By comparing the rarity score of all NFTs in a collection, we can rank the rarity of each NFT, with 1 being the rarest.
\end{itemize}

In this paper, I use the Bored Ape Yacht Club (BAYC) collection as an example. The Bored Ape Yacht Club is a well-known collection of 10,000 unique Bored Ape NFTs. The collection was created by Yuga Labs in April 2021. BAYC has a total trading volume of 1.4 million ETH (\$2.2 billion USD) as of December 2023. BAYC is favored by many celebrities and has a large community of collectors.

\subsection{Data sources}
I obtained data from two sources: Dune Analytics and OpenSea.

\subsubsection{Dune Analytics}
Dune Analytics is a platform for querying public databases from the Ethereum blockchain. For each collection, I query the NFT trades database and gather information on all NFT transactions from issuance to Sep 30, 2023.

\subsubsection{OpenSea}
OpenSea is the largest NFT marketplace focusing on NFTs based on Ethereum and Ethereum's Layer-2 ecosystem. The data from OpenSea's public API include data on NFT collections. In particular, I extracted data on the traits and rarity of each unique NFT in the collections. The set of available traits varies by collection, but generally include color, background, face/expression, and accessories.

\subsection{Data description}
As of September 30, 2023, there has been 43,383 BAYC trades recorded in the Dune \texttt{nft.trades} database. It is worth noting that trades conducted outside of marketplaces, such as through private over-the-counter agreements, are not recorded, however those agreements occur not as frequently. Here I outline the features and target variables used in this project. The features are divided into three categories: market, traits \& rarity, and last trade. The features are described in Table \ref{tab: feature_columns}.
\begin{table}[H]
    \centering
    \caption{Feature Columns Description (Bored Ape Yacht Club)}
    \begin{tabular}{|l|l|l|}
    \hline
    \textbf{Feature Category} & \textbf{Feature} & \textbf{Description} \\ \hline

    \multirow{4}{*}{Market} & volume\_eth\_lag1 & daily market volume (ETH) \\ \cline{2-3} 
    & price\_p5\_eth\_lag1 & daily 5-percentile price (ETH) \\ \cline{2-3} 
    & price\_max\_eth\_lag1 & daily highest price (ETH) \\ \cline{2-3} 
    & price\_min\_eth\_lag1 & daily minimum (floor) price (ETH)\\ \hline

    \multirow{8}{*}{Traits \& Rarity} & rarity\_rank & rarity rank measured by OpenRarity \\ \cline{2-3} 
    & Background\_count & number of items with the same \textit{Background} trait \\ \cline{2-3} 
    & Mouth\_count & number of items with the same \textit{Mouth} trait \\ \cline{2-3} 
    & Eyes\_count & number of items with the same \textit{Eyes} trait\\ \cline{2-3} 
    & Fur\_count & number of items with the same \textit{Fur} trait \\ \cline{2-3}
    & Clothes\_count & number of items with the same \textit{Clothes} trait \\ \cline{2-3} 
    & Earring\_count & number of items with the same \textit{Earring} trait \\ \hline

    \multirow{2}{*}{Last Trade} & last\_trade\_timediff & time since the last time this item was traded \\ \cline{2-3} 
    & last\_trade\_price & price at the last time this item was traded\\ \hline
    \end{tabular}
    
    \label{tab: feature_columns}
\end{table}

Note that all market data (volume and prices) are at daily frequency and lagged by 1 day. The traits and rarity data are static and do not change over time. The last trade data is dynamic and changes with each trade. For each trait, I use the count of items in the collection which shares the same trait (e.g. \texttt{Fur="tan"} corresponds to \texttt{Fur\_count=626}), with lower counts signaling higher rarity of the specific trait. The outcome variable is the trade price of the NFT (in ETH).

\subsection{Exploratory analysis}

\subsubsection{Price history}
\begin{figure}[H]
    \caption{Trade Price by Date}
    \includegraphics[width=\textwidth]{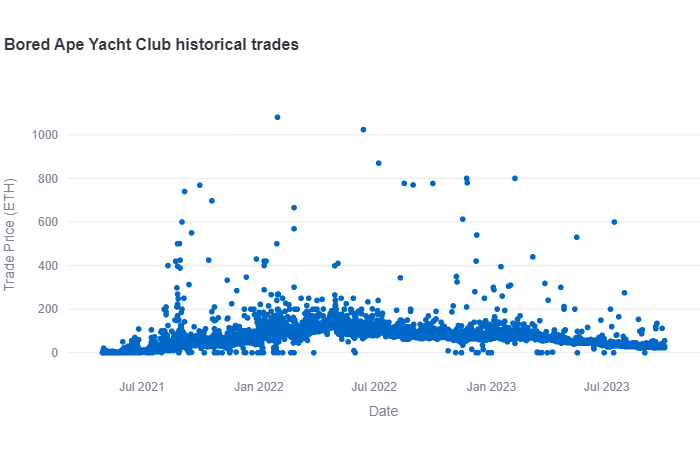}
    \label{fig: price_date}
\end{figure}

\begin{figure}[H]
    \caption{Trade Price by Rarity Rank}
    \includegraphics[width=\textwidth]{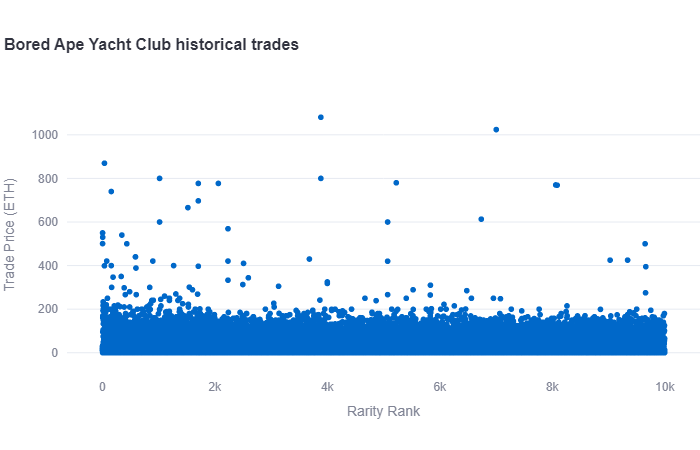}
    \label{fig: price_rarity}
\end{figure}

Figure \ref{fig: price_date} plots the price history of all BAYC trades since April 2021. Figure \ref{fig: price_rarity} plots the price history of BAYC trades by rarity rank. The plots show that the price of BAYC trades has increased significantly since the collection was launched. The price of BAYC trades is positively correlated with rarity rank, with the rarest NFTs trading at a premium.

\subsubsection{Principal component analysis}
\begin{figure}[H]
    \caption{PCA(3) plot of NFT prices (Bored Ape Yacht Club)}
    \includegraphics[width=\textwidth]{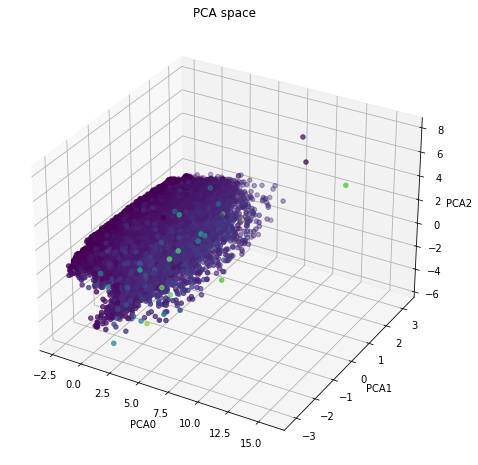}
    \label{fig: pca}
\end{figure}

Figure \ref{fig: pca} plots the first three principal components of the traits and rarity features. The plot shows that the traits and rarity features are not linearly separable, which motivates the use of non-linear models such as tree-based models and deep learning models.   

\section{Methodology}
\label{sec: method}
\subsection{Feature selection and benchmark models}
To find the set of features with the highest predictive power, I define three feature sets and examine the predictive outcome use linear and tree-based models.

Approximately 20\% of the trades dataset feature first-time trades, which lack the last\_trade features. This implies a skewed distribution of trading frequency among BAYCs. Therefore one major goal of this step is to examine the choice of including the last\_trade variables at the cost of dropping that portion of training data. 

I define three feature sets, each containing a combination of feature categories:
\begin{itemize}
    \item Set 1: market, traits \& rarity, last trade
    \item Set 2: market, traits \& rarity
    \item Set 3: traits \& rarity, last trade
\end{itemize}

For each feature set, I fit three types of linear models (OLS, Lasso, Ridge) and two types of tree-based ensemble models (random forest, gradient boosting), selecting the feature set that yields the best outcome metric.

As a robustness check, I use the same model-feature set pairs to predict an alternative outcome variable: the log-transformed trade price of each trade.

\subsection{Deep learning models}
I define a class of convolutional neural network (CNN) and use hyperparater tuning to find the architecture that results in minimum loss. The choice for CNN is motivated by the cross-sectional nature of the data. Convolutional layers are designed to capture high-order interations between the traits features and market movements. The CNN model is tuned using random search over the hyperparameters described in Section \ref{sec: hyperparam_tuning}.

\section{Results}
\label{sec: results}
\subsection{Benchmarking}
Benchmarking results using linear and tree-based models are reported in Table \ref{tab:benchmark}
\begin{table}[H]
    \centering
    \caption{MSE Loss Metric for Benchmark Models and Explanatory Variable Sets}
    \label{tab:benchmark}

    \begin{subtable}{\linewidth}
    \centering
    \caption{Y = Trade price (ETH)}
    \begin{tabular}{lccc}
    \hline
    Model & X1 & X2 & X3 \\ 
    \hline
    OLS & 353.02 & 319.01 & 762.74 \\
    Lasso & 374.11 & 327.36 & 772.64 \\
    Ridge & 353.02 & 319.01 & 762.74 \\
    Random Forest & 154.97 & 200.96 & 254.85 \\
    Gradient Boosting & 143.20 & 200.50 & 262.19 \\
    \hline
    \end{tabular}
    \end{subtable}%
    \bigskip
    
    \begin{subtable}{\linewidth}
    \centering
    \caption{Y = log (Trade price (ETH))}
    \begin{tabular}{lccc}
    \hline
    Model & X1 & X2 & X3 \\ 
    \hline
    OLS & 0.97 & 1.63 & 1.52 \\
    Lasso & 2.07 & 2.73 & 2.73 \\
    Ridge & 0.97 & 1.63 & 1.52 \\
    Random Forest & 0.20 & 0.43 & 0.27 \\
    Gradient Boosting & 0.22 & 0.43 & 0.28 \\
    \hline
    \end{tabular}
    \end{subtable}%
    \bigskip
    
    \begin{subtable}{\linewidth}
    \centering
    \caption{Y = Trade price (USD)}
    \begin{tabular}{lccc}
    \hline
    Model & X1 & X2 & X3 \\ 
    \hline
    OLS & 5.01e+09 & 4.37e+09 & 8.29e+09 \\
    Lasso & 5.01e+09 & 4.37e+09 & 8.29e+09 \\
    Ridge & 5.01e+09 & 4.37e+09 & 8.29e+09 \\
    Random Forest & 1.29e+09 & 1.27e+09 & 2.91e+09 \\
    Gradient Boosting & 2.03e+09 & 2.29e+09 & 3.23e+09 \\
    \hline
    \end{tabular}
    \end{subtable}
\end{table}
    
\subsection{Deep Learning Models}

\subsubsection{Model architecture}
\begin{figure}[H]
    \centering
    \caption{Example CNN Architecture. The actual architecture is tuned using \texttt{keras\_tuner} package.}
    \label{fig:cnn_arch}
    \includegraphics[width=\textwidth]{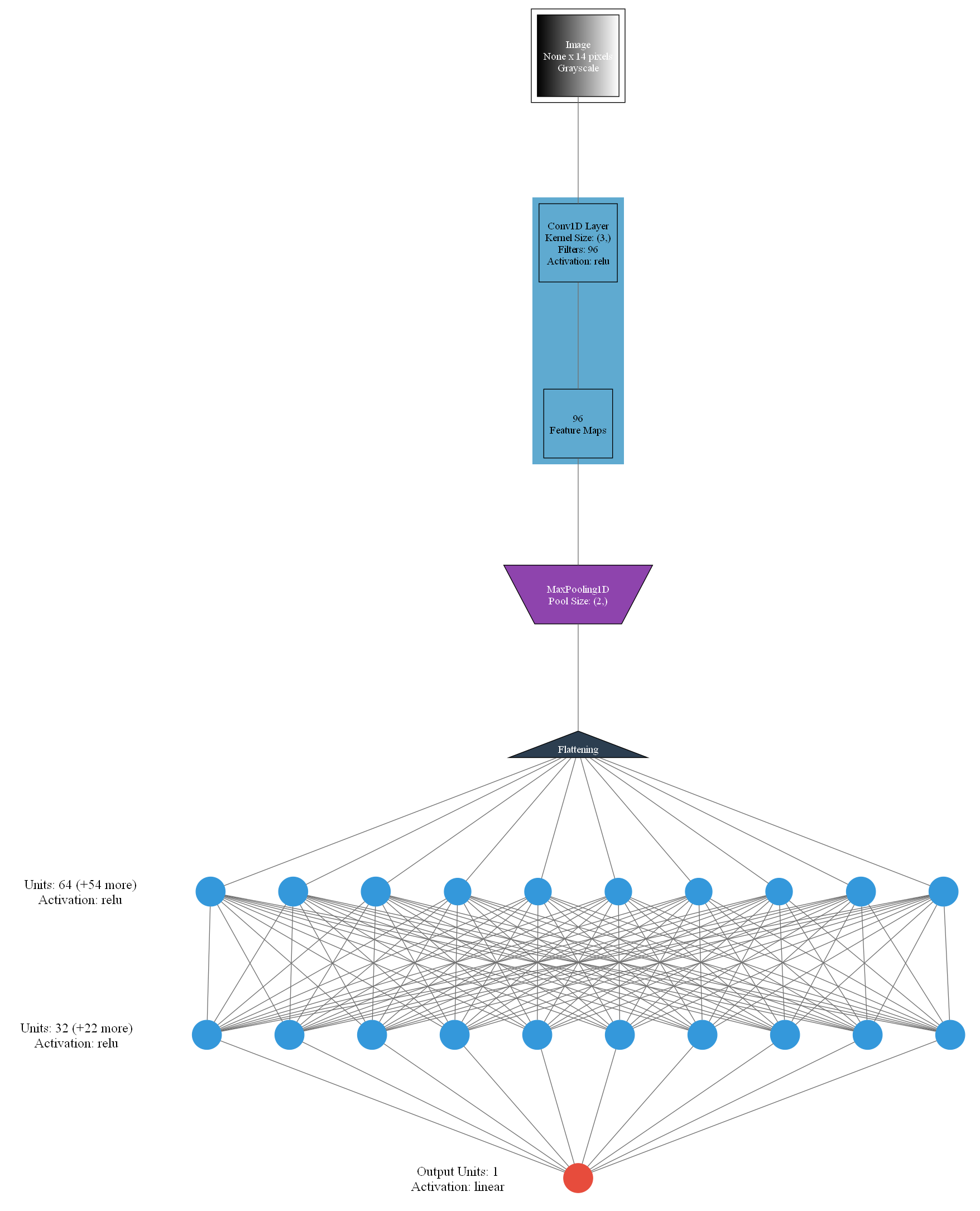}
\end{figure}

The deep learning models are trained using the \texttt{tensorflow.keras} package in \texttt{Python 3.9} under an \texttt{WSL-2 Ubuntu} operating system. Figure \ref{fig:cnn_arch} illustrates the model architecture, without the optional dropout layer.

\subsubsection{Hyperparameter tuning}
\label{sec: hyperparam_tuning}

The tunable hyperparameters include: number of filters and kernel size in the convolutional layer, number of units in the first dense layer, a boolean to include the dropout layer, and the learning rate for Adam optimizer. The feature importance and correlation metrics are computed post-hoc using the weights \& biases package. 

\begin{table}[H]
    \centering
    \caption{List of Hyperparameters and Importance with Respect to Validation Loss.}
    \begin{tabular}{l|c|c}
    \hline
    \textbf{Config Parameter} & \textbf{Importance} & \textbf{Correlation} \\ \hline
    kernel size             & .427            & -.102  \\ 
    learning rate           & .356            & .055  \\ 
    filters                  & .100            & .008  \\ 
    units                    & .087            & -.091  \\ 
    dropout                  & .030            & -.098  \\ \hline
    \end{tabular}
    \label{tab:cnn_hyperparam}
\end{table}

\begin{figure}[H]
    \centering
    \caption{CNN Training and Validation Loss by Epoch. Each line represents a trial with different hyperparameters}
    \label{fig:cnn_loss}

    \begin{subfigure}{\textwidth}
        \caption{Training loss by epoch}
        \includegraphics[width=\textwidth]{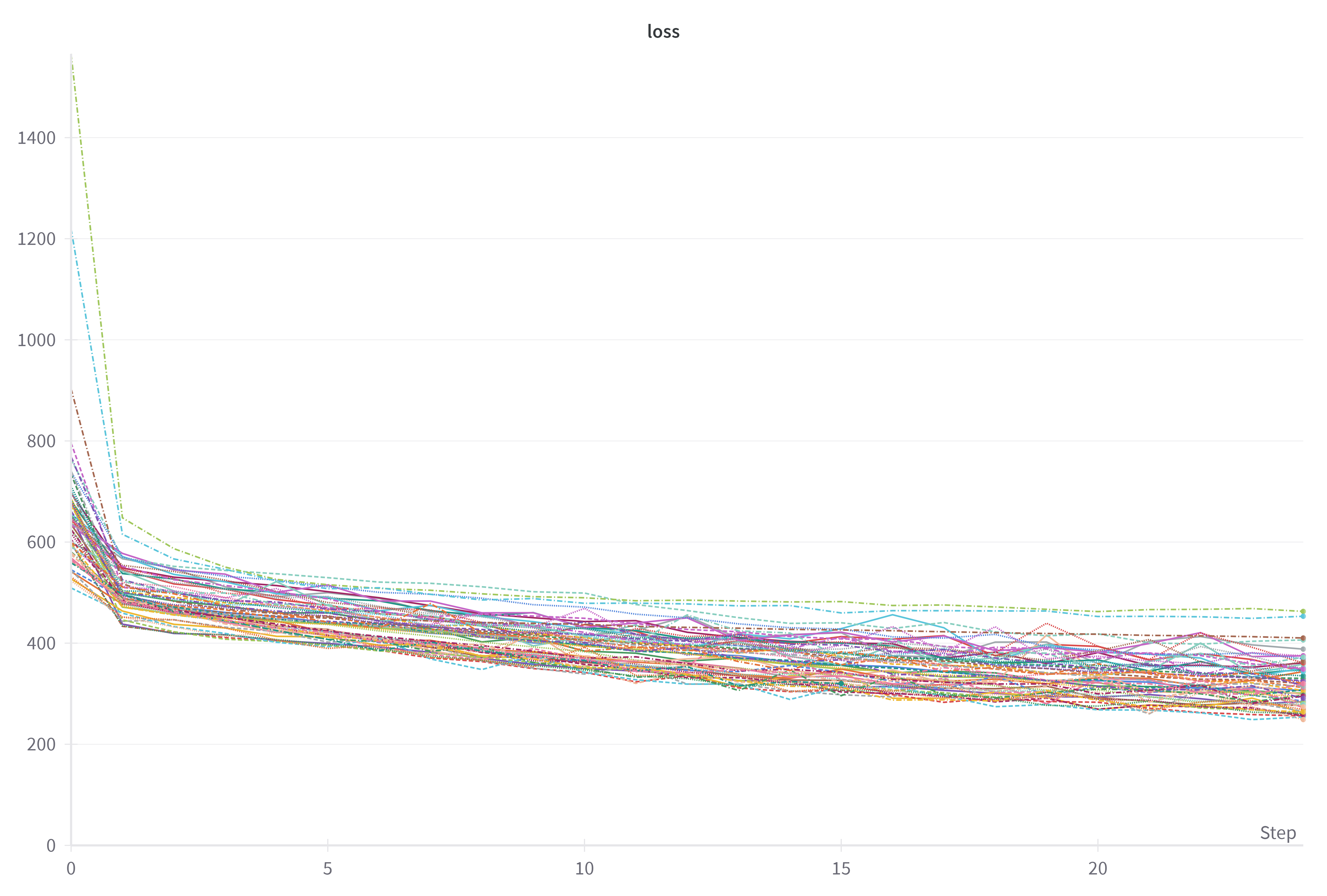}
    \end{subfigure}
    \begin{subfigure}{\textwidth}
        \caption{Validation loss by epoch}
        \includegraphics[width=\textwidth]{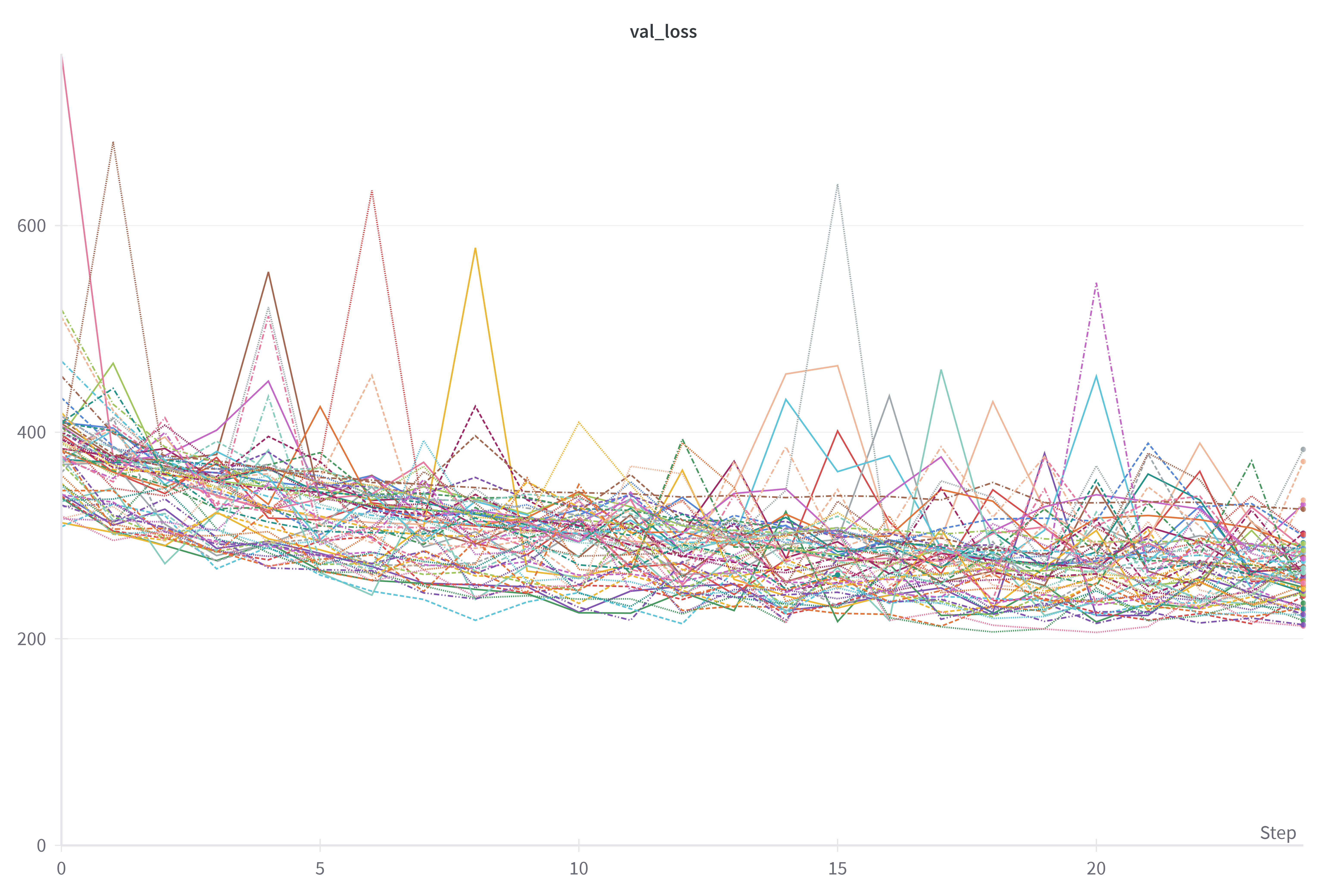}
    \end{subfigure}
\end{figure}

\begin{figure}
    \caption{CNN Hyperparameters Searched using \texttt{keras\_tuner.RandomSearch} method}
    \label{fig:cnn_tuning}
    \includegraphics[width=\textwidth]{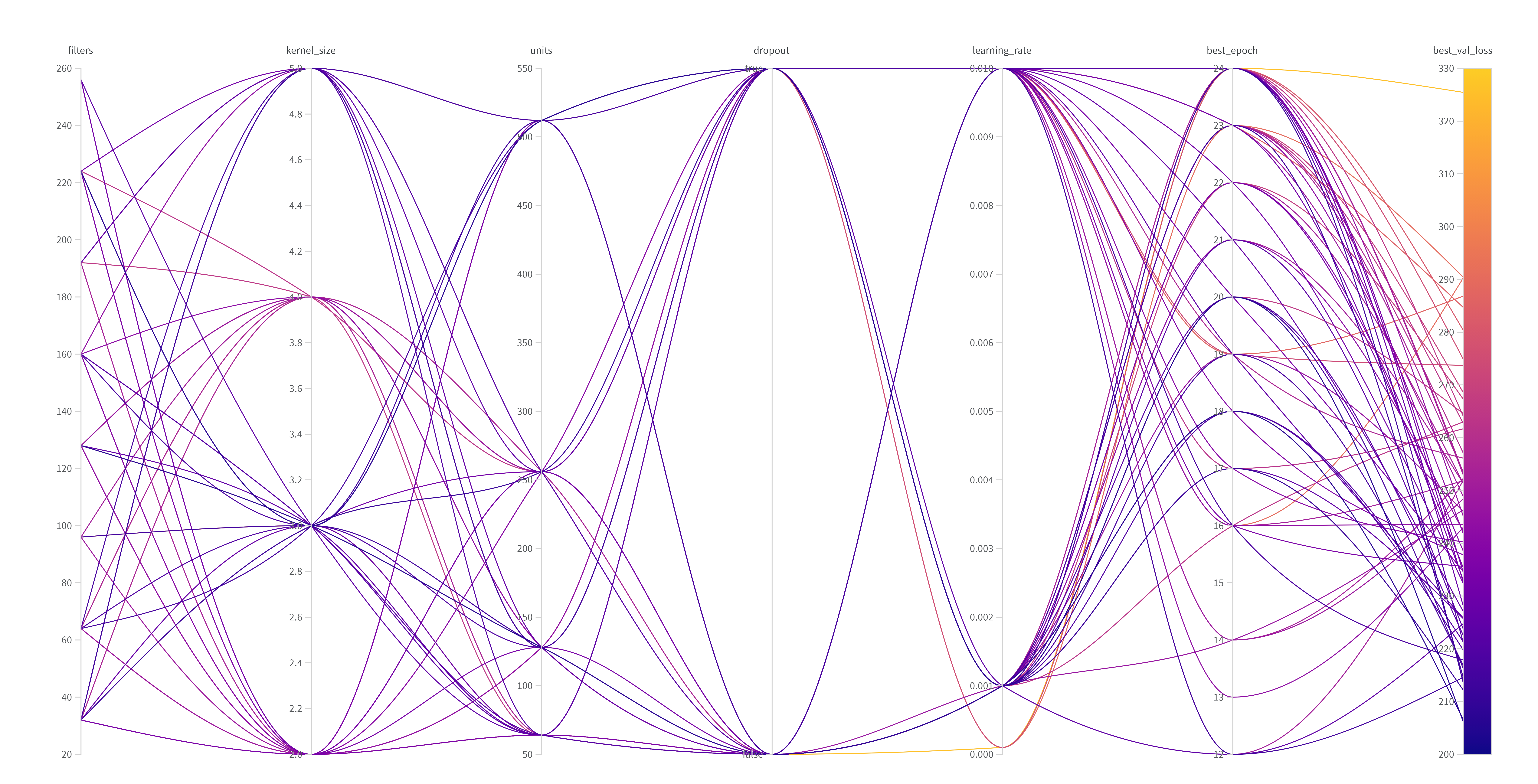}
\end{figure}

I tune these hyperparameters using the \texttt{RandomSearch} method in \texttt{keras\_tuner} package.  During training, the number of epochs is capped at 25, with a batch size of 32. The training and validation loss by epoch is shown in Figure \ref{fig:cnn_loss}. The searched hyperparameters and resulting validation loss (MSE) are shown in Figure \ref{fig:cnn_tuning}.

\section{Application}
\label{sec: application}
In this section, I propose an application framework for this model using zero-knowledge machine learning (zkML) and discuss its potential use cases in the context of decentralized applications (DApps). 

The goal of the framework is to provide a dynamic and trustless valuation model for NFTs while can be easily integrated with onchain NFT marketplaces and decentralized finance (DeFi) applications. 

\subsection{Dynamic valuation}
Dynamic valuation empowers the use of NFTs as collateral for DeFi applications, a recent trend in the NFT market. Borrowing against NFTs is currently limited and risky due to the highly volatile market (e.g. ``pump-and-dumps") and the lack of a reliable valuation model for volatile NFT prices. Most existing implementations rely on human input such as user-defined offers, direct negotiations, and Dutch auctions. The incorporation of a reliable valuation model can greatly improve the efficiency of these applications, and allows for dynamic adjustment of borrowing rates and auto-liquidations to protect lenders.

Valuation for an NFT is conducted as follows. For each NFT in a collection (e.g. BAYC \#8817), a collection-specific model is able to output a real-time valuation in ETH based on 1. the last trade price and time since last trade 2. the traits and rarity of the NFT 3. the current market conditions (e.g. volume, price floor, etc.) Due to the dynamic nature of the model, the valuation of each NFT can be updated in real-time as new trades occur and market conditions change.

\subsection{Trustless valuation}
In the traditional banking and loan sector, valuation models are considered top level enterprise secrets and are never published. Hence all users must rely on trust in the enterprise to fairly value their collateralized property, such as a house. 

In decentalized applications, however, such level of trust is difficult to establish. This proposed framework solves the issue of trustless valuation by using zero-knowledge machine learning. zkML allows for verifiable proof that a price valuation is indeed calculated by the model, instead of e.g. a manipulated price oracle. Meanwhile, the proof does not reveal the parameters of the model itself. This solves the problem of trustless valuation while preventing exploitation. Recent works by \citet{liu2021zkcnn,zhang2020zero} have demonstrated the feasibility of zkML in ensemble and deep learning models. 

In addition, the design choice of using tabular data for all model inputs generates great benefits for fully onchain applications, since storing and transferring image data onchain is costly.

\subsection{Technical implementation}
The technical implementation of this framework consists of three components: a data pipeline, a family of pre-trained prediction models, and a smart contract. The data pipeline is responsible for collecting and processing data from the Ethereum blockchain. The prediction model is responsible for generating the real-time valuation of each NFT. The smart contract is responsible for emitting the valuation of each NFT and providing zero-knowledge proofs onchain.

\section{Conclusion}
\label{sec: conclusion}

This research provides insights into valuation of Non-Fungible Tokens (NFTs), particularly in the context of dynamic markets. I provide an innovative approach to the dynamic valuation of NFTs, using deep learning models and a rich cross-sectional dataset from the Ethereum blockchain and OpenSea. Using the Bored Ape Yacht Club NFTs as a demonstration, the model integrates historical trades, market trends, and traits/rarity features to achieve high accuracy in price prediction. 

In real-world applications, the incorporation of zero-knowledge machine learning (zkML) can further enhances the model's utility, opening up new possibilities in decentralized finance (DeFi) applications. This work not only contributes to the field of digital asset valuation but also sets a foundation for future research in dynamic and trustless NFT pricing within the evolving landscape of blockchain technologies and decentralized markets.


\newpage
\nocite{*}
\printbibliography

\newpage
\appendix

\section{Rarity Score and Rarity Rank}
\label{app: rarity}
I use OpenSea's OpenRarity standard to calculate the rarity of each NFT within its collection. The rarity score is defined as follows:\\
For an NFT $x$ with traits $i\dots n$, its rarity score is
\[R(x) = \frac{I(x)}{\mathbb{E}[I(x)]}, \text{where } I(x)=\sum_{i=1}^n -\log_2P(trait_i).\] 


\end{document}